\documentclass[epj]{svjour}
\usepackage{amsmath,amssymb,amsfonts,graphicx,cite,algorithmic,algorithm,multirow}
\graphicspath{{graphics/}}
\makeatletter
\ifx\input@path\@undefined
\def\input@path{{graphics/}}
\else
\g@addto@macro\input@path{{graphics/}}
\fi
\makeatother

\preprint{DESY 13-237\\ MCnet-13-21}

\title{Summing Large-N Towers in Colour Flow Evolution}

\author{Simon Pl\"atzer}

\institute{DESY, Notkestrasse 85, D-22607 Hamburg, Germany}

\date{\today}

\abstract{ We consider soft gluon evolution in the colour flow
  basis. We give explicit expressions for the colour structure of the
  (one-loop) soft anomalous dimension matrix for an arbitrary number
  of partons, and show how the successive exponentiation of classes of
  large-$N$ contributions can be achieved to provide a systematic
  expansion of the evolution in terms of colour supressed
  contributions.
   }

\begin{document}

\maketitle


\section{Introduction}

In order to reliably interpret current and upcoming measurements at
the LHC, precise QCD predictions for multi-jet final states are
indispensable. These include both fixed-order calculations, as well as
their combination with analytic resummation and/or parton shower event
generators, {\it e.g.}
\cite{Sjostrand:2007gs,Bahr:2008pv,Gleisberg:2003xi}, to sum leading
contributions of QCD corrections to all orders, such as to arrive at a
realistic final state modelling.  Fixed-order calculations at leading
and next-to-leading order in the strong coupling are by now highly
automated, and frameworks to automatically resum a large class of
observables have been pioneered as well \cite{Banfi:2003je}. The
combination of NLO QCD corrections with event generators
\cite{Frixione:2002ik,
  Nason:2004rx,Nagy:2005aa,Platzer:2011bc,Hoeche:2011fd,Hartgring:2013jma}
is an established research area, and first steps towards combining
analytic resummation and event generators have been undertaken
\cite{Alioli:2012fc}.

The efficient treatment of QCD colour structures is central to both
fixed-order and resummed perturbation theory. Particularly the use of
the colour flow basis has led to tremendously efficient
implementations of tree-level amplitudes
\cite{Maltoni:2002mq,Duhr:2006iq,Gleisberg:2008fv}, which can be used
both for leading order calculations, as well as one-loop corrections
within the context of recent methods requiring only loop integrand
evaluation (see \cite{Reuschle:2013qna} for the exact treatment of the
colour flow basis in the one-loop case). This colour basis is closely
linked to determining initial conditions for parton showering. After
evolving a partonic system through successive parton shower emissions,
while keeping track of the colour structures (in the large-$N$ limit),
colour flows also constitute the initial condition to hadronization
models; this includes the dynamics of how multiple partonic
scatterings are linked to hadronization. Colour reconnection models,
such as those described in \cite{Sandhoff:2005jh,Gieseke:2012ft}, are
exchanging colour between primordial hadronic configurations like
strings or clusters, and have proven to be of utmost phenomenological
relevance in the description of minimum bias and underlying event
data.

Despite its relevance to event generators, the colour flow basis has
typically not been considered in analytic resummation, most probably
for the reason of being not the most simple or minimal basis. While
recent work has focused on obtaining minimal (and even orthogonal)
colour bases \cite{Keppeler:2012ih}, an intuitive connection to the
physical picture is hard to maintain in such approaches. It is until
now an open question, whether amplitudes can be evaluated in a
similarly efficient way in such bases. Also, in analytic resummation,
a matching to a fixed-order calculation is usually mandatory and the
use of colour flow bases could allow to use the full power of
automated matrix element generators within this context. Understanding
soft gluon evolution in the colour flow basis thus seems to be a
highly relevant problem to address, which can also shed light on
colour reconnection models, being so far based on rather simple
phenomenological reasoning.

The purpose of the present work is to study soft gluon evolution in
the colour flow basis. While for a fixed, small number of partons the
exponentiation of the soft gluon anomalous dimension matrix can be
performed either analytically or numerically, the case for a large
number of partons is rapidly becoming intractable. This limitation
thus prevents insight into the soft gluon dynamics of
high-multiplicity systems relevant to both improved\hfill parton\\ shower
algorithms \cite{Schofield:2011zi,Platzer:2012qg} as well as colour
reconnection models. We will derive the general structure of the soft
anomalous dimension matrix in the colour flow basis for an arbitrary
number of partons, and tackle its exponentiation by successive
summation of large-$N$ powers in a regime where the kinematic
coefficients $\gamma$ are of comparable size to the inverse of the
number of colours, $\gamma N \sim 1$, leading to a computationally
much more simple problem than the full exponentiation. This strategy
can well be applied to a large number of partons in an efficient way.

This paper is organized as follows: In section~\ref{section:anomdim}
we set our notation and present the general form of the soft gluon
anomalous dimension. In section~\ref{sections:towers} we derive its
exponentiation and show how subsequent towers of large-$N$
contributions can be summed in a systematic
way. Section~\ref{sections:numerics} is devoted to a few numerical
studies of testing the accuracy of these approximations in a simple
setting of QCD $2\to 2$ scattering, while
section~\ref{sections:outlook} presents an outlook on possible future
applications before arriving at conclusions in
section~\ref{sections:conclusions}. A number of appendices is devoted
to calculational details and for reference formulae to achieve what we
will later call a next-to-next-to-next-to-leading colour (N$^3$LC)
resummation.

\section{Notation and Soft Anomalous Dimensions}
\label{section:anomdim}

We consider the soft-gluon evolution of an amplitude $|{\cal
  M}_n\rangle$ involving $n$ coloured legs, either in the fundamental
or adjoint representation of $\text{SU}(N)$, with in general $N$
colour charges. The amplitude is a vector in both colour and spin
space, though we shall here mainly be interested in the colour
structure, decomposing the amplitude into a colour basis
$\{|\sigma\rangle\}$,
\begin{equation}
|{\cal M}_n\rangle = \sum_\sigma {\cal M}_{n,\sigma} |\sigma\rangle \ .
\end{equation}
We assume that all momenta of the amplitude are taken to be outgoing,
and will order the fundamental and adjoint representation legs
successively as 
$$
\alpha=1_{{\mathbf N}},2_{\bar{{\mathbf N}}},...,(n_q-1)_{{\mathbf N}},n_{q,\bar{{\mathbf N}}},
(n_q+1)_{{\mathbf A}},...,(n_q+n_g)_{{\mathbf A}}
$$ for the case of $n_q$ fundamental and anti-fundamental, and $n_g$
adjoint representation legs.  We will consider soft gluon evolution of
the amplitude,
\begin{equation}
|{\cal M}'_n\rangle = e^{\mathbf \Gamma} |{\cal M}_n\rangle \ ,
\end{equation}
with the soft anomalous dimension
\begin{equation}
{\mathbf \Gamma} = \sum_{\alpha\ne \beta} \Gamma^{\alpha\beta}\ {\mathbf T}_\alpha\cdot{\mathbf T}_\beta \ ,
\end{equation}
in terms of the usual colour charge products ${\mathbf
  T}_\alpha\cdot{\mathbf T}_\beta$. Though sometimes basis independent
results can be obtained for the soft gluon evolution, {\it e.g.}
\cite{Forshaw:2008cq}, one in general sticks to a particular basis of
colour structures in order to obtain a matrix representation of
${\mathbf \Gamma}$ such that the exponentiation can be performed.

\begin{figure}
\begin{center}
\includegraphics[scale=0.5]{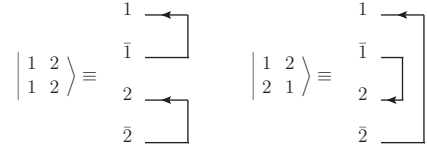}
\end{center}
\caption{\label{figures:basis}An illustration of the colour basis
  chosen for the case of two colour flows. Connected lines correspond
  to Kronecker-$\delta$ symbols in the space of (anti-)fundamental
  representation indices.}
\end{figure}

\begin{figure}
\begin{center}
\includegraphics[scale=0.5]{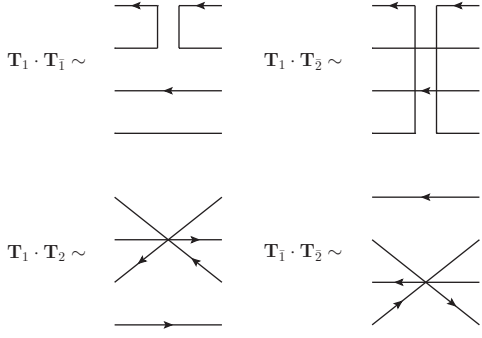}
\end{center}
\caption{\label{figures:correlators}Illustration of the non-diagonal
  contributions to colour charge products acting on colour flow basis
  tensors. Note that the `singlet' operators are entirely equivalent
  to the `swapping' ones.}
\end{figure}

We shall here consider the {\it colour flow basis}, by translating all
colour indices into indices transforming either in the fundamental
(${\mathbf N}$) or the anti-fundamental ($\bar{{\mathbf N}}$)
representation. For a thorough derivation of this paradigm, including
a list of Feynman rules and their application to fixed-order
calculation, see for example \cite{Maltoni:2002mq}.  Translating the
labelling of physical legs, $\alpha$, to a labelling of corresponding
colour and anti-colour `legs',
\begin{eqnarray}\nonumber
k & \leftrightarrow & \alpha = k_{\mathbf{N}} \\\nonumber
\overline{k-1} & \leftrightarrow & \alpha = k_{\bar{\mathbf{N}}}
\\\nonumber
\left.\begin{array}{c}
\overline{k-n_q/2}\\
k-n_q/2
\end{array}\right\}& \leftrightarrow & \alpha = k_{\mathbf{A}}
\end{eqnarray}
we are able to label the basis tensors in the colour flow basis by
permutations of the anti-colour indices relative to the colour indices,
\begin{equation}
|\sigma\rangle = \left|\begin{array}{ccc} 1 &  ... & m\\ \sigma(1) & ... & \sigma(m)\end{array} \right\rangle
= \delta^{i_1}_{i_{\overline{\sigma(1)}}} \cdots \delta^{i_m}_{i_{\overline{\sigma(m)}}} \ ,
\end{equation}
where $m=n_q/2+n_g$\footnote{Notice that we do not impose a limitation
  to colour structures as appearing for tree-level
  calculations. Indeed, the gluon exchange will generate all possible
  structures starting from only tree level ones.}.  A pictorial
representation of these basis tensors is given in
figure~\ref{figures:basis}.
The colour charges (note that ${\mathbf T}_\alpha\cdot {\mathbf
  T}_\beta = {\mathbf T}_\beta\cdot {\mathbf T}_\alpha$) translate as
(obvious cases relating colour and anticolour are not shown):
\begin{align}\nonumber
{\mathbf T}_\alpha\cdot {\mathbf T}_\beta = 
{\mathbf T}_k\cdot {\mathbf T}_l &\quad & \alpha = k_{\mathbf{N}}, \beta = l_{\mathbf{N}} \\\nonumber
{\mathbf T}_\alpha\cdot {\mathbf T}_\beta = 
{\mathbf T}_k\cdot {\mathbf T}_{\bar{l}} &\quad & \alpha = k_{\mathbf{N}}, \beta = (l+1)_{\bar{\mathbf{N}}} \\\nonumber
{\mathbf T}_\alpha\cdot {\mathbf T}_\beta = 
{\mathbf T}_k\cdot {\mathbf T}_{\bar{l}} + {\mathbf T}_{k}\cdot {\mathbf T}_{l}
&\quad & \alpha = k_{\mathbf{N}}, \beta = (l+n_q/2)_{\mathbf{A}}\\\nonumber
{\mathbf T}_\alpha\cdot {\mathbf T}_\beta = 
{\mathbf T}_{k}\cdot {\mathbf T}_{\bar{l}} + {\mathbf T}_{k}\cdot {\mathbf T}_{l} &\quad&
\alpha = (k+n_q/2)_{\mathbf{A}},\\\label{eqs:chargeTranslation}
+{\mathbf T}_{l}\cdot {\mathbf T}_{\bar{k}} + {\mathbf T}_{\bar{k}}\cdot {\mathbf T}_{\bar{l}}
&\quad & \beta = (l+n_q/2)_{\mathbf{A}}
\end{align}
and the colour flow charge products are expressed
as\footnote{$(t^a)^i {}_j (t^a)^k {}_l = \frac{1}{2}\left(\delta^i
  {}_l \delta^k {}_j - (1/N) \delta^i {}_j \delta^k {}_l\right)$}
\begin{equation}
{\mathbf T}_i\cdot {\mathbf T}_j =
\frac{1}{2}\left(\delta^{i'}_j \delta^{j'}_i - \frac{1}{N} \delta^{i'}_i \delta^{j'}_j\right)
\end{equation}
for a system of two ${\mathbf N}$ (and similarly for a system of two
$\bar{{\mathbf N}}$) legs, and by
\begin{equation}
{\mathbf T}_i\cdot {\mathbf T}_{\bar{j}} =
-\frac{1}{2}\left(\delta^{i'} {}_{\bar{j}'} \delta^{\bar{j}} {}_{i} - \frac{1}{N} \delta^{i'} {}_i \delta^{\bar{j}} {}_{\bar{j}'}\right)
\end{equation}
for a ${\mathbf N}\bar{{\mathbf N}}$ correlation\footnote{Note that
  appropriate crossing signs have to be included when considering
  incoming quarks, {\it i.e.}, a factor of -1 for each correlator
  involving an incoming quark or anti-quark as long as the anomalous
  dimension coefficients and amplitudes are evaluated in the physical
  regime.}. Hence the anomalous dimension reads
\begin{equation}
{\mathbf \Gamma} = 
\sum_{i<j} ( \gamma_{ij} {\mathbf T}_i\cdot {\mathbf T}_j +
\gamma_{\bar{i}\bar{j}} {\mathbf T}_{\bar{i}}\cdot {\mathbf T}_{\bar{j}} ) +
\sum_{i, j} \gamma_{i\bar{j}} {\mathbf T}_i\cdot {\mathbf T}_{\bar{j}}\ ,
\end{equation}
where the form of the $\gamma$ can be inferred from
eq.~\ref{eqs:chargeTranslation}, {\it e.g.}\footnote{Note that we did
  not assume $\Gamma^{\alpha\beta} = \Gamma^{\beta\alpha}$ in the
  first place, as may be due to inclusion of recoil effects or further
  contributions along the lines of dipole subtraction terms
  \cite{Catani:1996vz}}
\begin{align}
\gamma_{kl} = \Gamma^{\alpha\beta} + \Gamma^{\beta\alpha}
& \quad & \alpha = k_{\mathbf{N}}, \beta = l_{\mathbf{N}} \\\nonumber
\gamma_{k\bar{l}} = \Gamma^{\alpha\beta} + \Gamma^{\beta\alpha}
& \quad & \alpha = k_{\mathbf{N}}, \beta = (l+1)_{\bar{\mathbf{N}}} \\\nonumber
\gamma_{k\bar{k}} = 0
& \quad & \alpha = \beta = (k+n_q/2)_{\mathbf{A}} \ .
\end{align}
Examples of the non-diagonal part of the colour correlators are given
in figure~\ref{figures:correlators}.
Since the colour flow charge products effectively describe one-gluon
exchange between two colour flow lines, the general form of the matrix
representation of ${\mathbf \Gamma}$ is straightforwardly found to be
given by\footnote{Our notation $[ |...| ]$ indicates that we refer to
  the matrix element with respect to the given representation of the
  amplitude as a complex vector, and not the quantity $\langle\tau |
  {\mathbf \Gamma}|\sigma\rangle$, which will only coincide with the
  former in an orthonormal basis, that being not the case for the
  colour flow basis considered here, as well as not for most other
  colour bases.}
\begin{equation}
[ \tau | {\mathbf \Gamma} | \sigma ] = \left(- N \Gamma_\sigma
+ \frac{1}{N} \rho \right) \delta_{\tau\sigma} + \Sigma_{\tau\sigma}\ ,
\end{equation}
where 
\begin{equation}
\Gamma_\sigma = \frac{1}{2}\sum_{i} \gamma_{i\overline{\sigma(i)}} \ ,
\end{equation}
\begin{equation}
\rho = \frac{1}{2}\left(\sum_{i,j}
\gamma_{i\bar{j}} -
\sum_{i<j}
(\gamma_{ij} +\gamma_{\bar{i}\bar{j}})
\right) \ ,
\end{equation}
while the off-diagonal elements are given by
\begin{equation}
\label{eqs:sigmadef}
\Sigma_{\tau\sigma} = 
\sum_{i,j}\Sigma_{ij\tau(i)\tau(j)} \delta_{\tau(i)\sigma(j)} \delta_{\tau(j)\sigma(i)}
\prod_{k\ne i,j} \delta_{\tau(k)\sigma(k)}
\end{equation}
with 
\begin{equation}
\Sigma_{ijkl} = \frac{1}{2}(\gamma_{ij}+\gamma_{\bar{k}\bar{l}}
-\gamma_{i\bar{k}}-\gamma_{j\bar{l}}) \ ,
\end{equation}
{\it i.e.} only non-vanishing when connecting two basis tensors which
do not differ by more than a transposition in the permutation
identifying these (note that the Kronecker $\delta$'s in
eq.~\ref{eqs:sigmadef} ensure exactly one transposition between $\tau$
and $\sigma$ and the sum consists of solely one term).

\section{Summation of Large-$N$ Towers}
\label{sections:towers}

Though the exponentiation of the soft anomalous dimension matrix is
possible either analytically or using standard numerical algorithms
for a fixed (small) number of external legs, a general expression
seems yet out of reach, due to the rapid growth of the dimension of
colour space with the number of partons. In this section, we will
consider successive approximations to the full exponentiation by
subsequently summing towers of large-$N$ contributions, $\gamma^n
N^{n-k}$.  To derive the form of the large-$N$ towers, let us start
from the structure of the soft anomalous dimension matrix,
\begin{equation}
{\mathbf \Gamma}
\equiv N \underline{\Gamma} + \underline{\Sigma} + \frac{1}{N} \rho \underline{1} \ ,
\end{equation}
where we choose an arbitrary ordering of the permutations to identify
these with the indices of the rows and columns of the matrix
representation, such that $\underline{\Gamma}={\rm
  diag}(\{-\Gamma_\sigma\})$, and
$\underline{\Sigma}=(\Sigma_{\rho\sigma})$ does not contain any
diagonal elements.

The $n$-th power of the matrix representation then takes the form
\begin{equation}
{\mathbf \Gamma}^n \equiv
\sum_{k=0}^n\sum_{l=k}^n \left(\begin{array}{c}n\\k\end{array}\right) N^{n-l-k}\rho^k \underline{\Sigma}_{n-k,n-l} \ ,
\end{equation}
where $\underline{\Sigma}_{n,l}$ originates from powers of
$N\underline{\Gamma} + \underline{\Sigma}$,
\begin{equation}
\label{eqs:sigmapower}
(N\underline{\Gamma} + \underline{\Sigma})^n = \sum_{l=0}^n N^l \underline{\Sigma}_{n,l} \ ,
\end{equation}
with matrix elements given by (see appendix~\ref{sections:sigmarecursion}):
\begin{multline}
\label{eqs:sigmaelements}
(\underline{\Sigma}_{n,l})_{\tau\sigma} =\\
(-1)^l\sum_{\sigma_0,...,\sigma_{n-l}} \delta_{\tau\sigma_0}\delta_{\sigma_{n-l}\sigma}
\left(\prod_{\alpha=0}^{n-l-1} \Sigma_{\sigma_{\alpha}\sigma_{\alpha+1}}\right)\ \times\\
Q_{n-l,l}\left(\{\sigma_0,...,\sigma_{n-l}\},\Gamma\right) \ .
\end{multline}
Here $\Gamma = \{\Gamma_\sigma\}$ and the details of the polynomials
$Q_{k,l}$ are discussed in appendix~\ref{sections:qpolys}. The
exponentiation of the anomalous dimension matrix is then given by
\begin{multline}
\label{eqs:expsum}
[\tau | e^{\mathbf \Gamma} |\sigma] =
\sum_{l=0}^\infty \frac{(-1)^l}{N^l}\ \times\\\sum_{k=0}^l
\frac{(-\rho)^k}{k!}\sum_{\sigma_0,...,\sigma_{l-k}} \delta_{\tau\sigma_0}\delta_{\sigma_{l-k}\sigma}
\left(\prod_{\alpha=0}^{l-k-1} \Sigma_{\sigma_{\alpha}\sigma_{\alpha+1}}\right)\ \times\\
R(\{\sigma_0,...,\sigma_{l-k}\},\{\Gamma_\sigma\})
\end{multline}
where $R$ is worked out in appendix~\ref{sections:qpolys}.

We are now in the position to define successive summation of large-$N$
contributions. Eq.~\ref{eqs:expsum} suggests to define successive
summations at (next-to)$^d$-leading colour (N$^d$LC) by truncating the
sum at $l=d$. Owing to the properties of the $R$ functions, we find
that this prescription amounts to summing (schematically), the
following contributions (lower order contributions always implied):
\begin{eqnarray} \nonumber
\text{at LC} & : & 1 + \gamma N + \gamma^2 N^2 + ... \\\nonumber
\text{at NLC} & : & \left(\gamma + \frac{\gamma}{N}\right) (1 + \gamma N + \gamma^2 N^2 + ...) \\\nonumber
\text{at NNLC} & : & \left(\gamma + \frac{\gamma}{N}\right)^2 (1 + \gamma N + \gamma^2 N^2 + ...) \ ,
\end{eqnarray}
{\it i.e.} we consider a regime in which $\gamma N = {\cal O}(1)$ to
require resummation, while $\gamma\sim 1/N$ and $\gamma/N\sim 1/N^2$
can still be considered small in comparison the the $N$ enhancement of
the ${\cal O}(1)$ towers being resummed.  We shall also consider the
case that we have (trivially) exponentiated all contributions stemming
from the $\rho$ contribution to the anomalous dimension matrix. This
resummation, which we will here refer to as N$^d$LC', is obtained by
only considering the $k=0$ terms in eq~\ref{eqs:expsum}, while
redefining the $\Gamma_\sigma$ appropriately, $\Gamma_\sigma' =
\Gamma_\sigma-\rho/N^2$. Then we sum towers of
$$
1 + \left(\gamma + \frac{\gamma}{N^2}\right) N + \left(\gamma + \frac{\gamma}{N^2}\right)^2 N^2 + ...
$$ with a prefactor of $(N\gamma + \gamma/N)^d$ at N$^d$LC'.  Explicit
expressions of the $R$ functions as required through N$^3$LC are given
in appendix~\ref{sections:rexplicit}.  Explicitly, at leading colour
(LC), we have
\begin{equation}
[\tau | e^{\mathbf \Gamma} |\sigma] = \delta_{\tau\sigma} e^{-N\Gamma_\sigma} + \text{NLC} \ ,
\end{equation}
whereas at next-to-leading colour (NLC), we have
\begin{multline}
[\tau|e^{\mathbf \Gamma}|\sigma] =
\delta_{\tau\sigma}e^{-N\Gamma_\sigma}\left(1 + \frac{\rho}{N}\right) - \\
\frac{1}{N}\Sigma_{\tau\sigma}\frac{e^{-N\Gamma_\tau} - e^{-N\Gamma_\sigma}}{\Gamma_\tau-\Gamma_\sigma}
+ \text{NNLC} \ .
\end{multline}
Note that the NLC summation is sufficient to recover the anomalous
dimension matrix upon a next-to-leading order expansion,
\begin{equation}
\left.[\tau|e^{\mathbf \Gamma}|\sigma]\right|_{\text{NLC}} = \delta_{\tau\sigma} + 
[\tau|{\mathbf \Gamma}|\sigma] + {\cal O}(\gamma^2) \ .
\end{equation}
Also note that the structure of the approximated exponentiations
reflects the same approximation to be applied to the scalar product
matrix of the basis tensors: The basis is orthogonal at LC, at NLC
only scalar products between tensors differing by at most a
transposition need to be considered (and there is no non-vanishing
matrix element of the exponentiated soft anomalous dimension
connecting other tensors to this order), and similar observations apply
to higher order summations.

As a first assessment on the accuracy of the procedure outlined above,
let us consider the case of the evolution of two colour flows. Here,
the soft anomalous dimension matrix in the basis
$\{|12\rangle,|21\rangle\}$ takes the form
\begin{equation}
{\mathbf \Gamma} \equiv \begin{pmatrix} -N \Gamma_{12} + \frac{1}{N}\rho & \Sigma_{1212} \\
\Sigma_{1221} & -N \Gamma_{21} + \frac{1}{N}\rho \end{pmatrix} \ ,
\end{equation}
and its exact exponentiation is given by
\begin{multline}
\label{eqs:exp2by2}
e^{\mathbf \Gamma} \equiv
\frac{e^{\frac{1}{N}\rho} e^{-\frac{N}{2}(\Gamma_{12}+\Gamma_{21})}}{\kappa} \ \times \\
\begin{pmatrix} -\Delta \sinh\frac{\kappa}{2} + \kappa \cosh\frac{\kappa}{2} & 2\Sigma_{1212}\sinh\frac{\kappa}{2} \\
2\Sigma_{1221}\sinh\frac{\kappa}{2} & \Delta \sinh\frac{\kappa}{2} + \kappa \cosh\frac{\kappa}{2} \end{pmatrix}
\end{multline}
where $\Delta=N(\Gamma_{12}-\Gamma_{21})$ and
$\kappa=\sqrt{\Delta^2+4\Sigma_{1212}\Sigma_{1221}}$.  Let us for the
moment assume that all $\gamma$ are real (though this is of course not
the general case); considering then a {\it phase space} region for
which $\Delta^2\gg 4\Sigma_{1212}\Sigma_{1221}$, $\kappa\sim
|\Delta|$, we recover the NLC' approximation, {\it i.e.}, there is a
phase space region where {\it purely kinematic reasons} give rise to a
NLC' expansion without having actually considered the very size of $N$
itself. Note that the different treatment of $\rho$, either absorbing
it into a redefinition of the $\Gamma_\sigma$, or treating it as
subleading itself, amounts -- for the case of $q\bar{q}$ singlet -- to
either keeping $C_F=(N^2-1)/(2N)$ exactly or doing a strict large-$N$
limit with $C_F\sim C_A/2$. An observation that these different
prescriptions account for the bulk of subleading-$N$ effects in a
colour-improved parton shower evolution \cite{Platzer:2012qg} has
already been made, though we are far from drawing an ultimate
conclusion here.

\section{Numerical Results}
\label{sections:numerics}

In this section we consider numerical results on summing subsequent
large-$N$ towers for the case of QCD $2\to 2$ scattering, $p_1,p_2\to
p_3,p_4$ with a simple assumption on the anomalous dimension matrix,
\begin{eqnarray}
\Gamma^{12} = \Gamma^{34} &=& 
\frac{\alpha_s}{4\pi}\left( \frac{1}{2} \ln^2 \frac{s}{\mu^2} - i\pi \ln \frac{s}{\mu^2}\right) \\ \nonumber
\Gamma^{13} = \Gamma^{24} &=& 
\frac{\alpha_s}{8\pi} \ln^2 \frac{|t|}{\mu^2} \\ \nonumber
\Gamma^{14} = \Gamma^{23} &=& 
\frac{\alpha_s}{8\pi} \ln^2 \frac{|u|}{\mu^2} 
\end{eqnarray}
in terms of standard Mandelstam variables $s,t,u$ and some resolution
scale $\mu$. This anomalous dimension corresponds to a jet veto in a
typical parton shower resolution variable, but otherwise should rather
be thought of as a generic example. We refer to
\cite{Kidonakis:1998nf} for a detailed discussion and note that a
colour flow approach for the quark-quark case has already been
considered in \cite{Sotiropoulos:1993rd}.  We will explicitly consider
the matrix elements of the exponentiated anomalous dimension. For the
case of processes involving four (anti-) quarks, we can directly
compare to the analytic result in eq.~\ref{eqs:exp2by2}, while for the
other cases we study the convergence of successive approximations
(though exact results could also be obtained in these cases). All
calculations have been carried out with the C++ library
\texttt{CVolver}, which is available on request from the author.

\begin{figure}
\begin{center}
\input{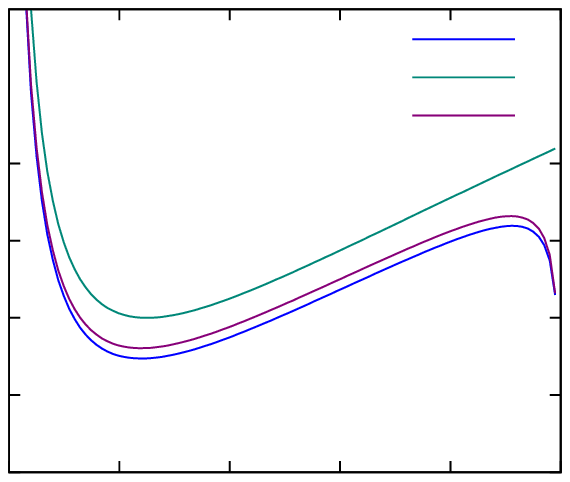}

\input{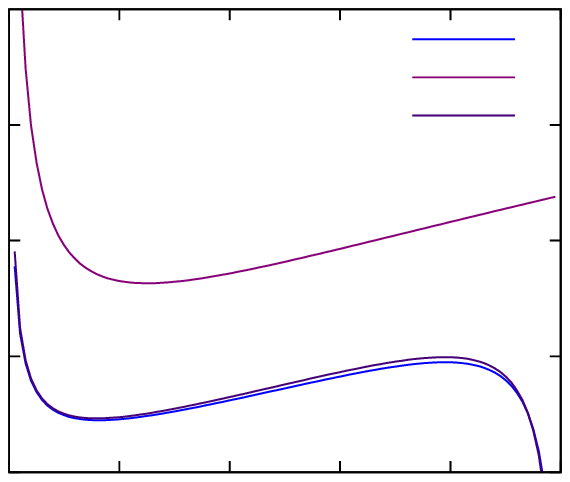}
\end{center}
\caption{\label{figures:diagonal}Real and imaginary parts of a
  diagonal evolution matrix element for quark-quark scattering at
  $s=100\ {\rm GeV}^2$, $\mu^2=25\ {\rm GeV}^2$ as a function of the
  momentum transfer $|t|$, comparing the exact results to various
  approximations. This matrix elements describes the amplitude to keep
  a $t$-channel colour flow $\sigma$.}
\end{figure}

\begin{figure}
\begin{center}
\input{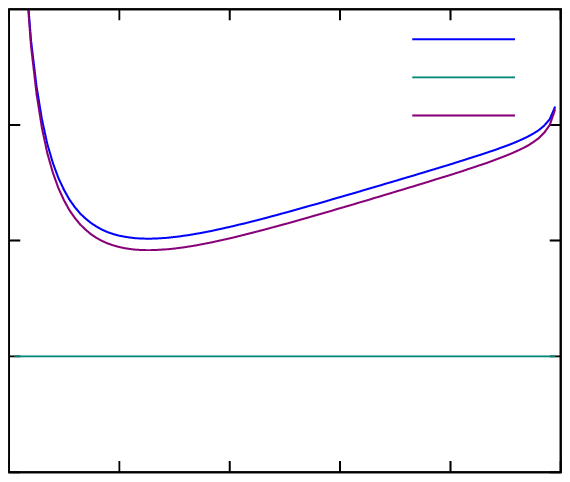}

\input{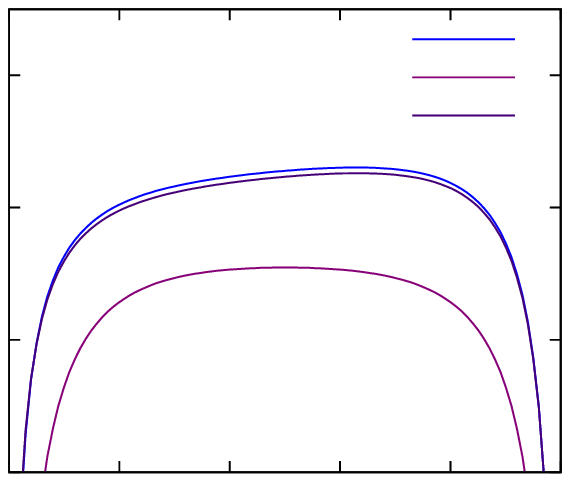}
\end{center}
\caption{\label{figures:offdiagonal}Same as
  figure~\ref{figures:diagonal} for an off-diagonal matrix
  element. The matrix element considered describes the transition from
  a $u$-channel colour flow $\tau$ to a $t$-channel one, $\sigma$.}
\end{figure}

\begin{figure}
\begin{center}
\input{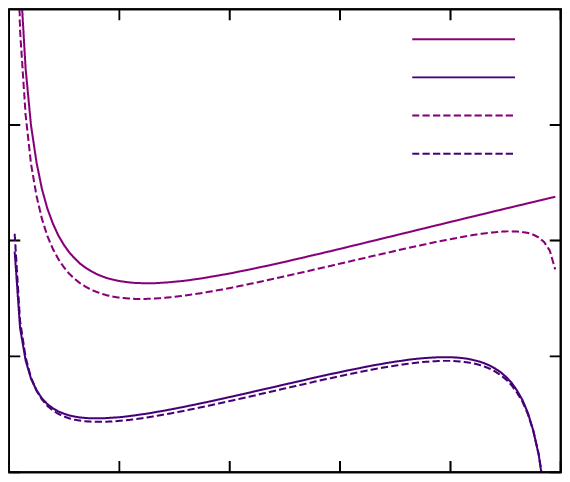}

\input{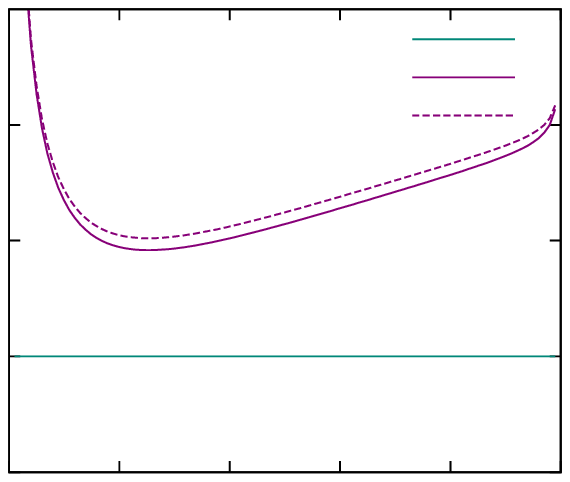}
\end{center}
\caption{\label{figures:primes}Comparison of the prime resummation
  prescription compared to the native one for the same parameters as
  used in figure~\ref{figures:diagonal}. Typically, a N$^2$LC'
  summation reaches a similar accuracy as a N$^3$LC one, both
  providing sub-permille agreement with the exact result.}
\end{figure}

For quark-quark scattering, we display numerical results for the real
and imaginary parts of the evolution matrix $e^{{\mathbf \Gamma}}$ in
figures~\ref{figures:diagonal} and \ref{figures:offdiagonal}.
Generally, we find that NLC summations are required to get a
reasonable approximation to the real part, while NNLC is required for
a similar description of the imaginary parts. At N$^3$LC we find a
sub-permille level agreement of the approximation with the exact
results.  In figure~\ref{figures:primes} we compare the difference
between the native summation and the prime prescription, which clearly
improves the approximation of the exact result leading to an accuracy
at N$^2$LC', which is comparable to the N$^3$LC calculation.

For the other configurations contributing to QCD $2\to 2$ scattering
we find a similar pattern of convergence through successive orders.
We note, however, that some of the matrix elements for processes with
more and more colour flows are non-zero starting only from a high
enough order.

\section{Outlook on Possible Applications}
\label{sections:outlook}

The work presented here is relevant to cases where soft gluon
evolution is a required ingredient for precise predictions, but not
feasible in exact form owing to a large number of external legs
present. This, in particular, applies to improved parton shower
algorithms but also to analytic resummation for observables of
multi-jet final states. Looking at the convergence of the N$^d$LC
expansions, which can easily be implemented in an algorithmic way, one
can gain confidence of providing a reliable resummed prediction at
some truncation of the exponentiation. As for the case of parton
showers, the colour flow basis, being itself ingredient to many highly
efficient matrix element generators, offers unique possibilities to
perform Monte Carlo sums over explicit colour structures or charges,
such that efficient algorithms in this case seem to be within reach.
The requirement to study soft gluon dynamics for a large number of
legs is as well at the heart of the dynamics behind non-global
logarithms \cite{Dasgupta:2001sh}, when considered to more than the
first order in which they appear, and beyond leading colour.  Another
application (which, in part, triggered the present work) is to gain
insight into the dynamics of colour reconnection models. A QCD
motivated and feasible colour reconnection model based on summing
large-$N$ towers is subject to ongoing work and will be presented
elsewhere.

Let us finally remark that N$^d$LC calculations in general do not
require matrix exponentiation and at most $d$ plain matrix
multiplications. Owing to the respective matrices being very
sparse\footnote{Note that this does not only apply to the colour flow
  basis, but similar observations have been made for other choices,
  {\it e.g.} \cite{Sjodahl:2009wx}}, this can be performed very
efficient. Indeed, one can imagine to perform a Monte Carlo summation
over colour structures by generating subsequent sequences of colour
flows to be considered. The number of possible sequences is very
limited given the fact that the $\Sigma$ matrices only contain
non-vanishing matrix elements for two colour flows which differ at
most by a transposition in the permutations labelling them.

\section{Conclusions}
\label{sections:conclusions}

In this paper we have investigated soft gluon evolution in the colour
flow basis, presenting the structure of the soft anomalous dimension
for any number of legs. We have then focused on systematic summation
of large-$N$ enhanced terms with the aim of providing successive
approximations to the exact exponentiation of the anomalous
dimension. We generally find a good convergence of these approximations
for a simple anomalous dimension in QCD $2\to 2$ scattering. The
present work can be used to perform soft gluon resummation for a large
number of external legs, where the full exponentiation is not feasible
anymore. It also forms the basis for improved parton shower evolution
and may shed light on the dynamics to be considered for colour
reconnection models.

Particularly in conjunction with matrix element generators, making use
of the colour flow basis, very efficient and highly automated
calculations can be performed owing to the algorithmic structure of
N$^d$LC approximations, including Monte Carlo sums over individual
colour structures. The C++ library \texttt{CVolver}
\cite{Platzer:CVolver}, which has been developed within this context
provides all required tools to do so.

\section*{Acknowledgments}

I am grateful to Malin Sj\"odahl and Mike Seymour for many valuable
discussions and comments on the work presented here.  This work has
been supported in part by the Helmholtz Alliance `Physics at the
Terascale'.

\appendix

\section{Anomalous Dimension Powers at $\rho =0$}
\label{sections:sigmarecursion}

In this appendix we consider powers of the non-trivial part of the
anomalous dimension, as decomposed in eq.~\ref{eqs:sigmapower}.
Introducing the boundary conditions $\underline{\Sigma}_{1,1}=
\underline{\Gamma}$, $\underline{\Sigma}_{1,0}= \underline{\Sigma}$
and $\underline{\Sigma}_{n,l}=0$ whenever $l<0$ or $l>n$ we find the
recursion
\begin{equation}
\underline{\Sigma}_{n,l} =
\underline{\Gamma}\ \underline{\Sigma}_{n-1,l-1} + 
\underline{\Sigma}\ \underline{\Sigma}_{n-1,l} \ ,
\end{equation}
which can be solved by
\begin{equation}
\underline{\Sigma}_{n,l} =
\sum_{m_0=0}^l \underline{\Gamma}^{m_0}\prod_{\alpha=1}^{n-l} \left(\sum_{m_\alpha = 0}^l \underline{\Sigma}\underline{\Gamma}^{m_\alpha}\right)
\delta_{\sum_{\beta=0}^{n-l}m_\beta,l} \ ,
\end{equation}
which satisfies all boundary conditions. Taking matrix elements of
this expression and inserting $(\underline{\Gamma})_{\tau\sigma} =
-\delta_{\tau\sigma} \Gamma_\sigma$, we arrive at the nested sum
expression given in equation~\ref{eqs:sigmaelements}.

\section{Summing $Q$-Polynomials}
\label{sections:qpolys}

Given a vector $\Gamma$ and a set of indices $\sigma =
\{\sigma_0,...,\sigma_k\}$ referring to elements in $\Gamma$ we
consider the following class of polynomials $Q_{l}$,
\begin{equation}
Q_{l}(\sigma,\Gamma) = 
\prod_{\alpha=0}^{\#\sigma-1} \left( \sum_{m_\alpha = 0}^l \Gamma_{\sigma_\alpha}^{m_\alpha}\right) \delta_{\sum_{\beta=0}^{\#\sigma-1} m_\beta,l} \ .
\end{equation}
Note that $Q_{l}(\{\sigma_0\},\Gamma) = \Gamma_{\sigma_0}^l$,
$Q_{0}(\sigma,\Gamma) = 1$ and $Q_{l}(\sigma,\Gamma) = 0$ for
$l<0$. Also note that $Q_l(\sigma,\Gamma)$ is independent of the order
of indices considered in $\sigma$. Let us first cover the case that
some of the indices in $\sigma$ are identical. Let $d_\alpha(\sigma)$
be the degeneracy of the index $\sigma_\alpha\in\sigma$, {\it i.e.}
$\sigma_\alpha$ occurs $d_\alpha(\sigma)+1$ times in $\sigma$. Also
let $\text{uniq}(\sigma)$ denote the set which is obtained by removing
all repeated occurrences of indices in $\sigma$. Using
\begin{equation}
Q_l(\underbrace{\sigma_0,...,\sigma_0}_{k+1\text{ times}},\Gamma) =
\frac{1}{k!} \frac{\partial^k}{\partial\Gamma_{\sigma_0}^k}\left(\Gamma_{\sigma_0}^k Q_l(\{\sigma_0\},\Gamma)\right)
\end{equation}
and the definition of $Q_l$ we then have
\begin{multline}
Q_l(\sigma,\Gamma) =\\ \left[\prod_{\alpha=0}^{\#\text{uniq}(\sigma)-1}
  \frac{1}{d_\alpha(\sigma)!}
  \frac{\partial^{d_\alpha(\sigma)}}{\partial\Gamma_{\sigma_\alpha}^{d_\alpha(\sigma)}}\Gamma_{\sigma_\alpha}^{d_\alpha(\sigma)}\right]
Q_l(\text{uniq}(\sigma),\Gamma) \ .
\end{multline}
If all indices are distinct, we have
\begin{multline}
Q_{l}(\text{uniq}(\sigma),\Gamma) =\\
\sum_{\alpha = 0}^{\#\text{uniq}(\sigma)-1} (\Gamma_{\sigma_\alpha})^{\#\text{uniq}(\sigma)+l-1}
\prod_{\beta=0, \beta\ne \alpha}^{\#\text{uniq}(\sigma)-1} \frac{1}{\Gamma_{\sigma_\alpha} - \Gamma_{\sigma_\beta}}
\end{multline}
which follows from using
\begin{equation}
\sum_{m=0}^{n} a^m b^{n-m} = \frac{a^{n+1}-b^{n+1}}{a-b}
\end{equation}
and the recursion
\begin{multline}
Q_{l}(\text{uniq}(\sigma),\Gamma) =\\
\frac{\Gamma_{\sigma_\alpha}}{\Gamma_{\sigma_\alpha} - \Gamma_{\sigma_\beta}}
Q_{l}(\text{uniq}(\sigma)\backslash \sigma_\beta,\Gamma)
+ (\alpha \leftrightarrow \beta) \ .
\end{multline}
We will especially need
\begin{multline}
R(\{\sigma_0,...,\sigma_{l-k}\},\Gamma) = \\
\sum_{n=l}^\infty \frac{(-N)^{n-k}}{(n-k)!} Q_{n-l}(\{\sigma_0,...,\sigma_{l-k}\},\Gamma)
\end{multline}
for $0 \le k \le l$. In this case,
\begin{equation}
R(\sigma,\Gamma) = 
\sum_{n=0}^\infty \frac{(-N)^{n}}{n!} Q_{n-\#\sigma+1}(\sigma,\Gamma)
\end{equation}
such that
\begin{multline}
R(\sigma,\Gamma) = \\ \left[\prod_{\alpha=0}^{\#\text{uniq}(\sigma)-1}
  \frac{1}{d_\alpha(\sigma)!}
  \frac{\partial^{d_\alpha(\sigma)}}{\partial^{d_\alpha(\sigma)}\Gamma_{\sigma_\alpha}}\Gamma_{\sigma_\alpha}^{d_\alpha(\sigma)}\right]
\tilde{R}(\sigma,\Gamma) \ ,
\end{multline}
where
\begin{multline}
\tilde{R}(\sigma,\Gamma)= \sum_{n=0}^\infty \frac{(-N)^{n}}{n!} Q_{n-\#\sigma+1}(\text{uniq}(\sigma),\Gamma) =\\
\sum_{\alpha = 0}^{\#\text{uniq}(\sigma)-1} (\Gamma_{\sigma_\alpha})^{\#\text{uniq}(\sigma)-\#\sigma}\ e^{-N\Gamma_{\sigma_\alpha}}\ \times\\
\prod_{\beta=0, \beta\ne \alpha}^{\#\text{uniq}(\sigma)-1} \frac{1}{\Gamma_{\sigma_\alpha} - \Gamma_{\sigma_\beta}}\ .
\end{multline}
Finally,
\begin{multline}
R(\sigma,\Gamma) = \left[\prod_{\alpha=0}^{\#\text{uniq}(\sigma)-1}
  \frac{1}{d_\alpha(\sigma)!}
  \frac{\partial^{d_\alpha(\sigma)}}{\partial^{d_\alpha(\sigma)}\Gamma_{\sigma_\alpha}}\right]\\ \sum_{\alpha
  = 0}^{\#\text{uniq}(\sigma)-1} e^{-N\Gamma_{\sigma_\alpha}}
\prod_{\beta=0, \beta\ne \alpha}^{\#\text{uniq}(\sigma)-1}
\frac{(\Gamma_{\sigma_\beta}/\Gamma_{\sigma_\alpha})^{d_\beta(\sigma)}}{\Gamma_{\sigma_\alpha}
  - \Gamma_{\sigma_\beta}}\ .
\end{multline}

\section{$R$ Functions through N$^3$LC}
\label{sections:rexplicit}

In this appendix we give explicit expressions for the $R$ functions
needed for summations through N$^3$LC. Note that the index order does
not matter. Also note that equality of some of the $\Gamma_\sigma$ is
equivalent to putting the respective indices to be equal.

\subsection{LC}

\begin{equation}
R(\{\sigma_0\},\Gamma) = e^{-N\Gamma_{\sigma_0}}
\end{equation}

\subsection{NLC}

\begin{equation}
R(\{\sigma_0,\sigma_1\},\Gamma) = \frac{e^{-N\Gamma_{\sigma_0}} - e^{-N\Gamma_{\sigma_1}}}{\Gamma_{\sigma_0}-\Gamma_{\sigma_1}}
\end{equation}

\begin{equation}
R(\{\sigma_0,\sigma_0\},\Gamma) = -N e^{-N\Gamma_{\sigma_0}}
\end{equation}

\subsection{NNLC}

\begin{multline}
R(\{\sigma_0,\sigma_1,\sigma_2\},\Gamma) = \\
\frac{e^{-N\Gamma_{\sigma_0}}}{(\Gamma_{\sigma_0}-\Gamma_{\sigma_1})(\Gamma_{\sigma_0}-\Gamma_{\sigma_2})}
+(0\leftrightarrow 1) + (0\leftrightarrow 2)
\end{multline}

\begin{multline}
R(\{\sigma_0,\sigma_0,\sigma_1\},\Gamma) = \\
-N \frac{e^{-N\Gamma_{\sigma_0}}}{\Gamma_{\sigma_0}-\Gamma_{\sigma_1}} +
\frac{e^{-N\Gamma_{\sigma_1}} - e^{-N\Gamma_{\sigma_0}}}{(\Gamma_{\sigma_1}-\Gamma_{\sigma_0})^2}
\end{multline}

\begin{equation}
R(\{\sigma_0,\sigma_0,\sigma_0\},\Gamma) = \frac{N^2}{2} e^{-N\Gamma_{\sigma_0}}
\end{equation}

\subsection{N$^3$LC}

\begin{multline}
R(\{\sigma_0,\sigma_1,\sigma_2,\sigma_3\},\Gamma) = \\
\frac{e^{-N\Gamma_{\sigma_0}}}{(\Gamma_{\sigma_0}-\Gamma_{\sigma_1})(\Gamma_{\sigma_0}-\Gamma_{\sigma_2})(\Gamma_{\sigma_0}-\Gamma_{\sigma_3})} +\\
(0\leftrightarrow 1) + (0\leftrightarrow 2) + (0\leftrightarrow 3)
\end{multline}

\begin{multline}
R(\{\sigma_0,\sigma_0,\sigma_1,\sigma_2\},\Gamma) = \\ -N \frac{e^{-N
    \Gamma_{\sigma_0}}}{(\Gamma_{\sigma_0}-\Gamma_{\sigma_1})(\Gamma_{\sigma_0}-\Gamma_{\sigma_2})}\ +\\ \frac{(\Gamma_{\sigma_1}
  + \Gamma_{\sigma_2} - 2\Gamma_{\sigma_0})\ e^{-N \Gamma_{\sigma_0}}}
{(\Gamma_{\sigma_0}-\Gamma_{\sigma_1})^2(\Gamma_{\sigma_0}-\Gamma_{\sigma_2})^2}\ +\\ \left(
\frac{e^{-N\Gamma_{\sigma_1}}}{(\Gamma_{\sigma_0}-\Gamma_{\sigma_1})^2(\Gamma_{\sigma_1}-\Gamma_{\sigma_2})}
+ (1\leftrightarrow 2)\right)
\end{multline}

\begin{multline}
R(\{\sigma_0,\sigma_0,\sigma_0,\sigma_1\},\Gamma) = 
\frac{N^2}{2}\frac{e^{-N\Gamma_{\sigma_0}}}{\Gamma_{\sigma_0}-\Gamma_{\sigma_1}}\ +\\
N \frac{e^{-N\Gamma_{\sigma_0}}}{(\Gamma_{\sigma_0}-\Gamma_{\sigma_1})^2} +
\frac{e^{-N\Gamma_{\sigma_0}}-e^{-N\Gamma_{\sigma_1}}}{(\Gamma_{\sigma_0}-\Gamma_{\sigma_1})^3}
\end{multline}

\begin{multline}
R(\{\sigma_0,\sigma_0,\sigma_1,\sigma_1\},\Gamma) =\\
-N\frac{e^{-N
    \Gamma_{\sigma_0}} + e^{-N
    \Gamma_{\sigma_1}}}{(\Gamma_{\sigma_0}-\Gamma_{\sigma_1})^2} -
2 \frac{e^{-N
    \Gamma_{\sigma_0}} - e^{-N
    \Gamma_{\sigma_1}}}{(\Gamma_{\sigma_0}-\Gamma_{\sigma_1})^3}
\end{multline}

\begin{equation}
R(\{\sigma_0,\sigma_0,\sigma_0,\sigma_0\},\Gamma) = -\frac{N^3}{6} e^{-N\Gamma_{\sigma_0}}
\end{equation}

\bibliography{largen-towers}

\end{document}